# Human-AI Chain-of-Thought, a collaborative workspace for sensemaking, visualization, steering, validation, and trust.

Nathan Conklin
Department of Computer Science
Virginia Tech
Blacksburg, VA, USA
nathan.conklin@vt.edu

Chris North
Department of Computer Science
Virginia Tech
Blacksburg, VA, USA
north@cs.vt.edu

## ABSTRACT

In long conversations with reasoning models, prompts, answers, assumptions, evidence, and intermediate reasoning are spread across interfaces, making them difficult to inspect or revise together. We propose Human-AI Chain-of-Thought, a persistent collaborative workspace in which users can inspect and revise the record of their reasoning with an AI. Knowledge Cards link each exchange to available reasoning artifacts, provenance, and reconstructed reasoning structures. Users inspect these through linked textual and graphical views. Proposed branch-based interactions let users annotate, challenge, endorse, and selectively revise the context used for subsequent generation. Multiscale navigation and consequence-guided attention are intended to help users inspect long reasoning histories. The observability requirements distinguish generated rationales, external evidence, mechanistic measurements, and activation-based readouts. We present a prototype and an evaluation agenda for comprehension, steering, and calibrated reliance. We argue that explainable AI should help users inspect and revise shared reasoning throughout a conversation, as well as explain individual answers.

## 1 Introduction

We propose a persistent Human-AI workspace for inspecting and revising shared reasoning. Users identify assumptions, inspect textual or computational evidence, challenge explanations, and choose what informs later exchanges. The interface links each choice to its continuation and states what each observation establishes. This approach to explainable AI (XAI) may support sensemaking and calibrated reliance.

Chain-of-thought (CoT) prompting can improve model performance through generated intermediate steps [57], but these explanations can omit influences on an answer [53]. Monitorability needs empirical testing for each model and task [31]. Human-AI Chain-of-Thought records prompts, responses, available reasoning artifacts, and human interventions. This external record does not capture either participant's complete cognition.

We contribute a workspace architecture and research agenda linking persistent Knowledge Cards, source-linked reasoning views, branch-based context steering, and observability requirements. The prototype supports views and navigation. Context edits affecting subsequent inference, advanced multiscale abstraction, and internal-signal overlays remain proposed extensions.

## 2 Background & Related Work

ReTrace structures reasoning traces for comprehension [17], ReasoningFlow reconstructs discourse graphs [32], and ReasonDiag supports error diagnosis [7]. Hippo enables reasoning-topic review and modification [45].

Clearer explanations need not improve decisions. Graphionale finds that graphical rationales affect calibration differently across task modalities [55]. Highlighted CoT can help users recognize correct answers but also make incorrect ones persuasive [40]. Fernandes et al. found higher subjective appraisal without performance gains in their tested conditions [19]. Sun et al. show that certainty framing changes reliance [52]. Readability, preference, correctness, and calibrated trust need separate evaluation.

Feng et al.'s Knowledge Card supplies knowledge through specialized language models [18]. AStar's thought cards capture reusable reasoning patterns [58]. Ferreira's Knowledge Cards organize bounded domain knowledge with provenance [20]. Ours bundles an exchange. Sensecape and Graphologue establish multilevel sensemaking and linked text-diagram interaction [51], [26]. ReasonGraph, SPROUT, and interactive explanation interfaces provide visualization and authoring precedents [33], [35], [61].

Building on trace visualization [17], structured reasoning representations [32], direct reasoning modification [45], and persistent goal tracking [10], we focus on correcting shared context across turns. The proposed record links original exchanges, typed evidence, inferred cross-turn dependencies, and branch-specific context decisions. Users can trace claim reuse, revise context within backend limits, and compare branches. The record would link each decision to the inspected passage, context edit, and new continuation.

## 3 Knowledge Cards as a shared context record

Each exchange forms $K_i = (T_i, R_i, G_i, H_i, P_i)$. The Thought Card $T_i$ holds the full human prompt and AI response. $R_i$ holds reasoning artifacts, including tool-separated segments where available. $G_i$ is the reconstructed reasoning graph. $H_i$ records human annotations and interventions; $P_i$ records provenance. Cards explicitly record when reasoning information is unavailable. The conversation links cards chronologically and through separately identified semantic dependencies (Figure 1).

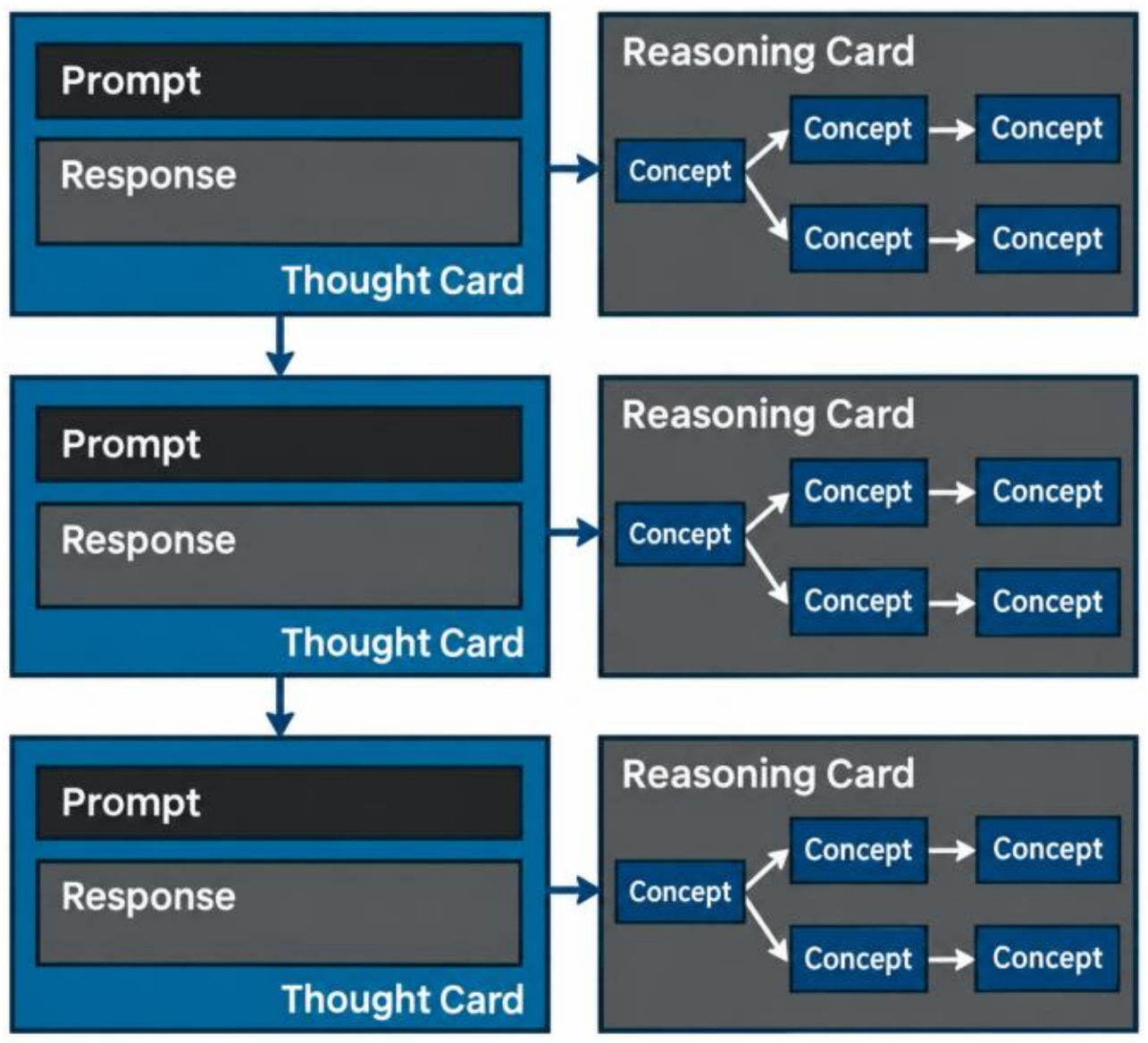

**Figure 1. The Knowledge Card chain (Human-AI Sensemaking) bundles prompt–response Thought Cards with associated Reasoning Cards (AI's CoT Reasoning).**

Each graph node retains its originating card and source span. Reasoning and summary representations remain distinct. A bundle may eventually contain Concept Graphs organizing recurring entities and ideas; their construction is deferred here. Content identifiers and preserved versions allow the human to revisit an earlier statement without treating a later reinterpretation as the original model output.

# 4 Workspace interaction and context steering

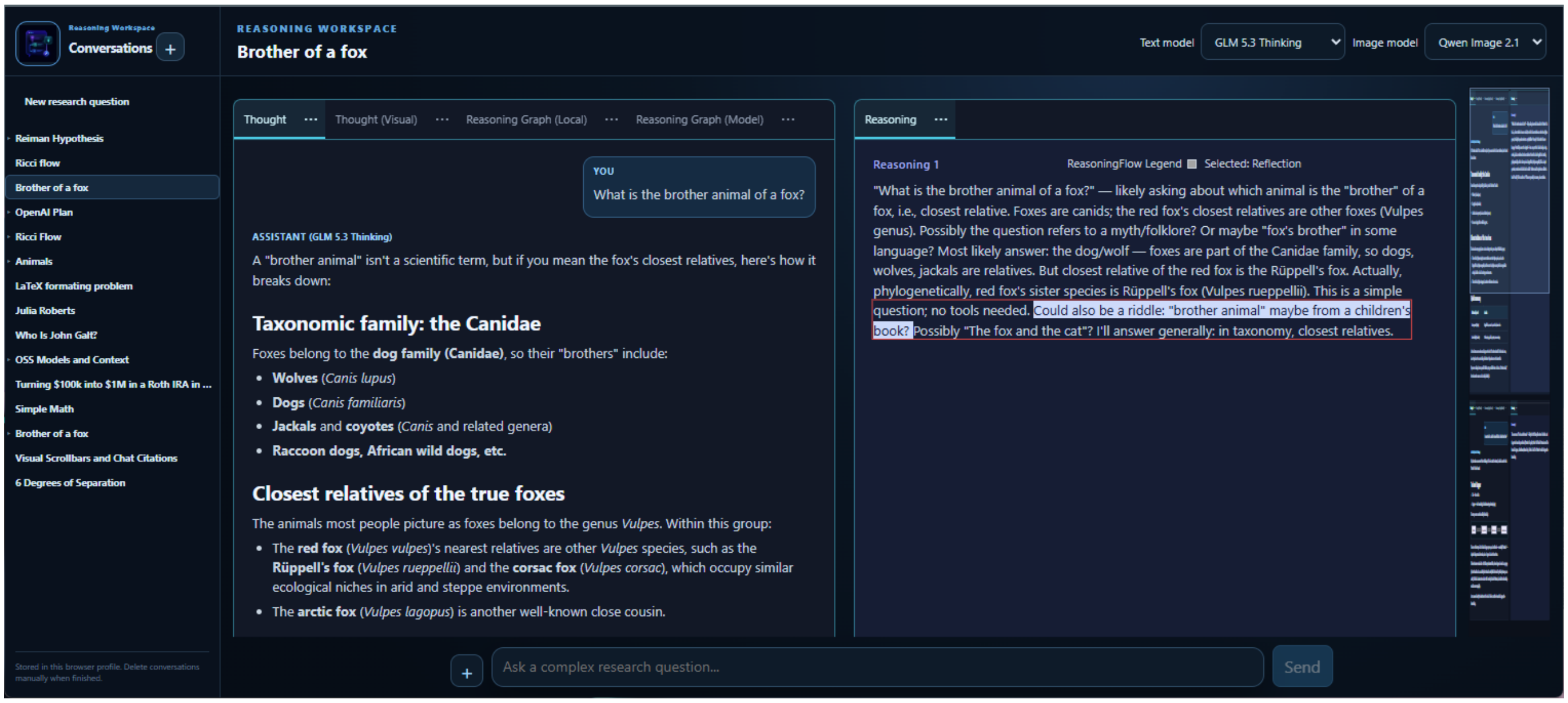

**Figure 2. The Reasoning Workspace displays prompt/response content alongside the associated reasoning artifacts. Users interact with the Knowledge or Reasoning to refine the context and steer the conversation.**

The prototype places prompts and responses beside reasoning text, with conversation navigation and an overview strip (Figure 2). Users can inspect reasoning while judging an answer. A linked graph shows dependencies they would otherwise reconstruct mentally. Selecting a passage for inspection is separate from changing future context.

## 4.1 A concrete steering episode

In the "Brother animal of a fox" example, the user asks for animal identification but asks it as an ambiguous question. The trace identifies the ambiguity and whether the user may be considering a children's book. The response and rationale differ, but this does not establish whether the model internally the various options within the reasoning.

In the interaction, the user annotates a selected passage: maybe from a children's book. They can endorse the habitat framing and exclude unwanted reasoning. Apply and branch interactions create an alternate continuation while preserving the original card and later exchanges on the original branch. The new branch records the edit, author, target spans, and submitted context.

## 4.2 Branching semantics

Display edits alone do not change model computation. For serialized branch context C[b] and human edit Δ, the constructor builds C[b′] = Assemble(C[b], Δ, capabilities) and requests a continuation. Removing a Knowledge Card excludes its prompt–response pair and reasoning. Removing a Reasoning Card or span retains the remaining exchange, subject to provider replay requirements. Revised prompts and answers become new branch artifacts. Edited answers are labeled as human revisions. Changes affect future context, not the original generation record.

Annotations and likes enter later requests as labeled human feedback. This neither implies training nor guarantees compliance. Branch construction must check retained answers, summaries, and retrieved material for restatements of excluded content. Removing a span does not establish removal elsewhere. The interface previews retained material and marks dependent artifacts for regeneration or review. The original record remains available for comparison. Unsupported reasoning edits require a fresh request using supported context and feedback.

The proposed intervention contract separates the historical record from approved context. For each operation, it records the source version, supported transformation, affected artifacts, resulting branch, and backend capabilities. This provenance makes the change auditable. The request must match the preview of retained, excluded, or substituted material. Contract checks are separate from instruction-following checks. Inferred cross-turn dependencies identify material for review. They neither prove causal influence nor guarantee complete semantic removal. Deleting displayed text does not establish its exclusion from future context when incompatible opaque continuity data remain. A compliant answer does not resolve that uncertainty.

## 5 Source-linked reasoning at multiple scales

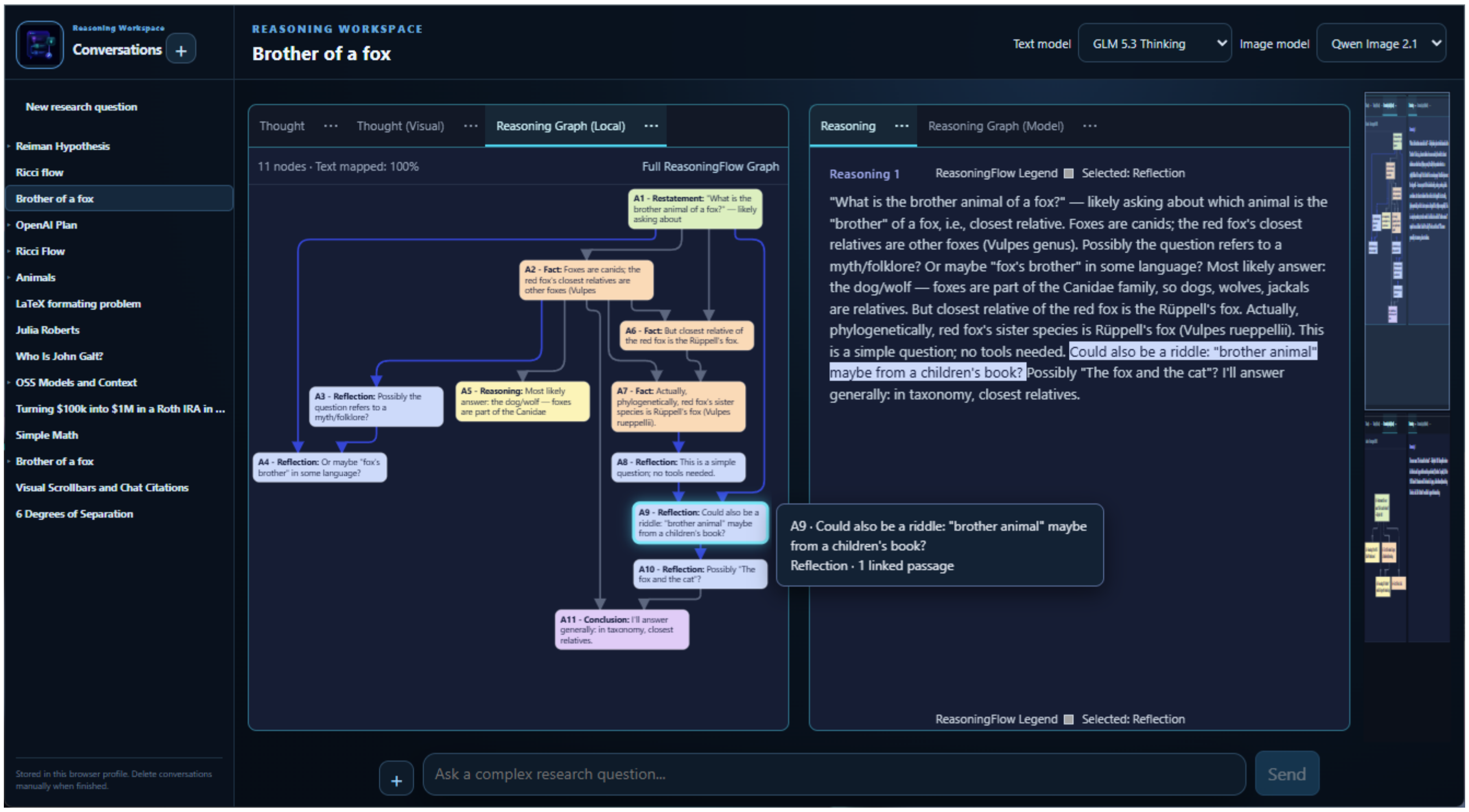

**Figure 3. Reasoning chains link a selected reasoning stages to source passages. The prototype supports graph navigation and linked highlighting. Nodes and edges reflect discourse in the reasoning trace (colored by ReasoningFlow palette [32]).**

Sentence sockets are used connect a selected reasoning stage to the words from which it was reconstructed (Figure 3). Selecting the source passage should also locate the associated graph node. Local graphs can follow ReasoningFlow's segmentation, role classification, and relation extraction [32]. Labels identify planning, assumptions, reflection, and other discourse functions. These are interpretations of emitted text. In particular, a discourse edge does not establish a neural causal pathway.

The proposed scaling design has four levels: conversation overview, card summary, grouped reasoning stages, and source passages. At the overview, users follow goals and revisions across turns. At the card level, they inspect conclusions and unresolved assumptions. Stage-level expansion reveals branching, verification, and abandoned approaches. Source-level inspection exposes the unabridged text. This applies overview and details-on-demand principles [48], [49] while retaining explicit links to provenance.

A summary preserves mappings to its constituent nodes, the number of collapsed steps, and unresolved disagreements. Automatic grouping must not hide a contradiction by replacing several historical claims with a single fluent statement. Cross-card links initially preserve separate occurrences and label reuse, revision, or contradiction; automatic merging could otherwise erase the history being inspected. Graph extraction failures should leave the original text usable. Prototype evaluation will measure abstraction fidelity as well as navigation speed.

The workspace can simplify the reasoning graph without losing its connection to the original text, while distinguishing a changed reasoning trace from a changed interpretation of that trace. The following rule is used for organizing and displaying the recorded reasoning. Formally, let $G = (V, E)$ be a typed discourse graph and $\pi$ assign each source node to a displayed group. The aggregate graph retains an edge between groups whenever a source edge crosses their boundary, with relation type and multiplicity preserved. Expansion restores the member nodes and their source offsets. A node record links a run, card version, source interval, extraction method, and discourse role. This proposed representation permits several abstractions of the same immutable trace. It also separates changes in the recorded reasoning from changes introduced by a newer extraction or grouping method.

## 5.1 Scaling attention as well as geometry

Long reasoning traces pose computational and usability problems. A graph may render successfully yet remain difficult to read due to the large number of nodes. We track token length, node count, edge density, dependency depth, and the number of conversation turns. ReasoningFlow's reports 247.7 thousand automatically annotated steps across 1,260 traces [32]. This is a large collection to inspect, but it does not imply that every trace has hundreds of dependency levels. Visual scaling is required to avoid overwhelming the user with overly complex graphs. The user interface reduces the amount exposed at once while retaining access to the complete recorded artifacts.

Beyond making reasoning histories easier to navigate, the workspace helps users decide which steps to inspect first. We propose prioritizing inspection by uncertainty and downstream consequences. For node v, an illustrative ordering rule is S(v) = U(v) × D(v), where U(v) is a task-specific uncertainty indicator and D(v) is the weighted consequence of conclusions depending on v. This is a design hypothesis, not an estimated probability of error or a validated measure of importance. Inferred dependencies make D(v) uncertain, and a low score must never prevent manual inspection. Critical contradictions and user-marked assumptions remain visible independently of the score.

The sources of U(v) must remain distinguishable: token surprise, disagreement among samples, verifier output, and human judgment measure different quantities. Generation probabilities alone did not provide the same benefit as edit-oriented highlighting in the code-completion study by Vasconcelos et al. [54]. We therefore propose testing uncertainty-only, consequence-only, and combined prioritization rather than assuming multiplication is optimal. The UI should explain why a passage is highlighted and indicate uncertainty in the graph extraction.

## 5.2 Interaction techniques

**Appendix A** identifies a fifteen-technique interaction design space and identifies their access requirements and prior art. Not all of these interactions are available yet, primarily because model APIs need to be enhanced to expose the necessary data to support the user interface interactions. Standardized extensions to the OpenAI API and model support are needed to help enable these future capabilities to improve interactions and expose model cognition [4].

An Assumption Ledger maintains claims that later exchanges reuse. Users can inspect a claim's source, endorse it, dispute it, or request clarification. OnGoal's persistent representation of conversational goals provides a related state-tracking precedent [10]. The proposed ledger distinguishes a human commitment from a model-inferred assumption; inferred contradictions require confirmation rather than silently rewriting session state.

A Reasoning Diff Between Turns aligns related passages or stages and shows additions, removals, and revisions. Users should be able to inspect the inferred alignment. Existing model-comparison interfaces offer useful visual precedents [50], [27], but this proposal concerns a developing conversation and its branches. Differences between two continuations cannot by themselves establish that an edit caused a change; generation variability and other context changes must be controlled in evaluation.

Provenance Descent lets a user move from an answer to a reasoning stage, its source passage, and supporting tool results or retrieved evidence. The descent may reach an instrumented internal measurement when available. Source citations, extracted dependencies, and activation readouts need distinct labels. When a lower layer is unavailable, the interface should say so and stop there, rather than generate a substitute that looks like a measurement.

A future Cross-Layer Agreement Gutter could place textual claims beside aligned verifier results and activation-conditioned readouts. The gutter would mark disagreements for inspection. Agreement would remain limited evidence because the instruments may share training data, model biases, or decoding assumptions.

## 6 Observability across the AI stack

The full application stack spans infrastructure, model execution, orchestration, retrieved evidence, persistent context, and human interaction (**Appendix B**). The workspace needs to associate observations from these layers with the same exchange. Tool success is evidence about an action; a citation identifies a source; a generated rationale states an approach; an activation records a selected internal state. The interface must align these observations and preserve their provenance without treating them as a single reasoning field.

OpenAI's hosted interface documents summaries, encrypted continuity, and reasoning.context for selecting reuse of available reasoning on supported models, while keeping the reasoning text opaque [42]. OpenAI-compatible Chat Completions interfaces for Kimi K3, DeepSeek-V4-Flash, and GLM-5.3 expose reasoning_content [38], [12], [59]. These names describe different capabilities: reasoning.context controls reuse; reasoning_content carries the reasoning text. Replay also differs. Kimi K3 requires the complete returned assistant message, including reasoning, to be preserved [38]. DeepSeek ignores prior reasoning_content when no tools parameter is supplied and incorporates it when tools are supplied [12]. An interoperable interface is needed to specify how evidence is captured, replayed, and changed by an intervention, as well as how its fields are named.

## 6.1 Beyond emitted reasoning

We propose interoperable, optional observability capabilities grounded in existing methods. Sparse autoencoders offer feature decompositions [11], and circuit tracing provides local attribution graphs in an approximate replacement model [1]. Activation patching and automated circuit discovery test intervention-sensitive components [9]. Interfaces need feature identity, layer and token alignment, reconstruction residuals, and the intervention metric before presenting such results as mechanistic evidence. Feature activation alone is insufficient to establish causal contribution.

Patchscopes, LatentQA, and Activation Oracles decode or answer questions about internal representations [24], [43], [28]. Introspection Adapters elicit learned behavioral dispositions [47], while Natural Language Autoencoders train a textual bottleneck using activation reconstruction [22]. The Jacobian lens provides vocabulary-oriented readouts based on local derivatives [25]. These methods motivate accessible explanations but measure different objects. A report about a modified model's learned behavior must not appear as a passive reading of an unmodified model's live state. CHIVE found no improvement from the interpretability techniques it tested in agents predicting counterfactual behavior [29]. We therefore propose transcript-only baselines to test whether the additional measurements improve decisions.

Required API capabilities include explicit availability reporting; typed evidence channels; stable run, model, and source identifiers; layer and position alignment; diagnostic-method versions; declared capture scope; and cost-aware, on-demand retrieval. Observation and intervention requests must remain distinguishable. Partial, redacted, or unsupported data should be reported explicitly. NNsight/NDIF, SGLang, and Neuronpedia demonstrate routes to instrumented access [21], [46], [39]. A standard must define the backend measurement represented by each client-side field.

Access to reasoning data also carries disclosure risks. Panfilov et al.'s Stealing Reasoning Traces from Proprietary LLM APIs documents risks involving proprietary traces and client-held encrypted reasoning [44]. We therefore recommend controlled research access that preserves provenance. This work could begin with instrumented open-weight deployments and controlled services, alongside proposals for capabilities that hosted providers can expose selectively. Evaluation must address three questions separately: whether useful evidence is available, whether it is interpreted faithfully, and how it affects human decisions.

# 7 Evaluation and Conclusion

Our team is conducting evaluation of the Human-AI Chain-of-Thought workspace to compare ordinary chat, chat with raw reasoning, and the Knowledge Card workspace. The conditions will match underlying conversations with both correct and incorrect advice. Participants will locate assumptions, trace dependencies, and decide whether to accept an answer. We plan to measure dependency comprehension and appropriate reliance separately, reporting acceptance of incorrect advice and rejection of correct advice alongside overall accuracy. Confidence judgments would support calibration analysis using Brier scores against answer correctness; time and workload would be secondary outcomes. Subjective trust alone is insufficient [55], [40], [19], [52], [6].

Steering will be evaluated by comparing targeted branch edits with ordinary follow-up prompts, using matched generation budgets and repeated continuations. We will record whether the requested change occurs, whether retained context reintroduces excluded content, and whether unrelated requirements are preserved. Scaling tests will vary length, density, and cross-turn dependencies to compare the four-level hierarchy and attention policies. Tests of internal observability will measure its incremental benefit over the same transcript, including shuffled or missing measurements where appropriate. Graph segmentation and dependencies will be validated against human annotations before attributing user effects to the visualization.

The key question is whether the linked record helps users correct a developing conversation, beyond making reasoning easier to read. We propose an editable-reasoning baseline using the same model, source artifacts, generation budget, and comparable editing operations. It would omit the workspace's explicit cross-turn links and intervention history. Tasks will require users to revise an early assumption that has been reused in later answers or summaries. We will measure identification of affected conclusions, recurrence of excluded content, preservation of unrelated requirements, and appropriate reliance. Removing dependency links and intervention provenance separately will test the contribution of each. If adequately powered, matched comparisons found negligible benefit, that result would weaken the architectural hypothesis even if users preferred the workspace.

Our study will recruit general chatbot users for bounded information and classification tasks and measure their expertise and familiarity with graphs. Randomized, counterbalanced tasks and held-out content can separate interface effects from learning. A pilot would inform the power analysis and task difficulty; we do not specify a sample size or expected effect at this stage. Failure cases include plausible but incorrect graph links, misleading highlights, unsupported replay, and readouts that sound more certain than their evidence permits.

Human-AI Chain-of-Thought is designed to let users follow how a conversation develops and revise the context used for subsequent exchanges. Its value will depend on whether users can better understand dependencies, make deliberate corrections, and judge when to rely on AI. The architecture specifies the data these interactions require, and the evaluation agenda sets out how to test their value.

## Appendix A Interaction technique catalog

This catalog describes candidate techniques for advanced user interface interactions that expose model cognition and provide improved user experiences to support sensemaking. Prior-art labels distinguish an existing interface precedent, a partial analogue, a method precedent, and a proposed integration. Several techniques use the same screen space, so the workspace should show only those suited to the current task.

| Technique | Description, access requirements, prior art, and limitations |
|---|---|
| **A1 Hesitation Ribbon** | Encode token surprise or distributional uncertainty alongside the text; selecting a segment reveals its measurement and available alternatives. GLTR visualizes token statistics, and HiLDe supports token-level intervention [23], [3]. Vendor probability displays are deployment-specific. Required access is token-aligned probabilities; the signal indicates properties of generation, not correctness or a human-like hesitation state. Status: existing interface precedents. |
| **A2 Cross-Layer Agreement Gutter** | Align emitted claims with independent checks and activation-conditioned reports, exposing disagreement through adjacent markers. Required access includes source alignment and method-specific readouts [24], [43], [28], [47], [22], [25]. How to normalize these different signals and account for correlated errors remains unresolved. Status: proposed integration; absence of a close precedent in this selected set does not establish novelty. |
| **A3 Knockout Probe** | Let a user exclude a cited source or selected context segment, rerun the same task, and inspect the difference. ContextCite and ERASER provide attribution and removal-based evaluation precedents [8], [13]. Repeated generations and matched controls are needed to distinguish context sensitivity from stochastic variation. Context removal does not constitute neural ablation. Status: method precedent, proposed in-chat interaction. |
| **A4 Sampled Path Distribution** | Show counts of independently sampled continuations grouped by semantic outcome. Self-consistency and semantic entropy motivate sampling and clustering [56], [16]; ChainForge and Loom provide interface precedents [5], [36]. Display sample count, sampling conditions, and clustering uncertainty. The samples are generated alternatives; they do not reveal hidden paths taken within a single run. Status: existing methods and separate-workspace precedents. |
| **A5 Divergence Point Marker** | Align sampled or branched traces and mark the earliest supported difference. Loom's branch points are a related structural precedent [36]. Semantic alignment can be ambiguous, and divergence location depends on the chosen representation. Required access is multiple source-preserved traces. Status: proposed extension of branching interfaces. |
| **A6 Assumption Ledger** | Maintain reusable assumptions with their source, author, status, and affected conclusions. OnGoal tracks conversational goals and Kiro specifications preserve explicit requirements [10], [30]. The proposed ledger would track disputed assumptions and the conclusions that depend on them. Required access is conversation history and human annotations. Status: partial interface precedents. |
| **A7 Open Questions Queue** | Preserve unresolved questions across exchanges and let users answer, defer, or dismiss them. Clarifying-question research and specification workflows provide precedents [2], [60], [30]. Distinguish user-entered questions from automatically inferred gaps. Status: partial precedents; the persistent queue is a proposed workspace integration. |
| **A8 Reasoning Diff Between Turns** | Compare related reasoning across exchanges or branches, retaining links to both source versions. LMdiff and LLM Comparator compare model outputs and provide visual comparison precedents [50], [27]. Cross-turn semantic alignment and attribution of changes remain research tasks. Required access is versioned traces; model-assisted matching may add inference cost. Status: proposed conversational adaptation. |
| **A9 Conversation Spine** | Provide a compact timeline of exchanges, goals, revisions, and intervention points. Sensecape and OnGoal establish conversation-level organization and navigation [51], [10]. The proposed spine is keyed to Knowledge Cards and branches. Required access is message order and stable card identifiers. Status: existing navigation precedents with a different persistent unit. |
| **A10 Frame Card** | Expose the intended task, constraints, and planned approach before generation, then let users revise them. Kiro's specification workflow is a practical precedent [30]. A reviewed frame becomes labeled context for the next request. Required access is an explicit planning stage; it should not imply that all subsequent internal computation will follow the plan. Status: existing workflow precedent. |
| **A11 Mid-Stream Intervention Window** | Offer intervention at meaningful generation or tool boundaries, with the continuation tied to the recorded prefix. HiLDe and AGDebugger provide token-level and agent-message precedents [3], [15]. OpenAI also documents mid-turn steering for a supported model and transport; queued input affects a continuation and does not undo emitted output or earlier actions [41]. Backend capabilities determine the permitted boundary. Status: existing methods and API support; proposed integration in this workspace. |
| **A12 Certainty Threshold Rendering** | Reduce initially displayed detail using explicitly identified confidence signals while keeping expand controls and unresolved contradictions visible. Progressive disclosure supplies a design precedent [49]. Essential qualifications must remain legible at low opacity, and high-confidence styling must not imply that a claim is true. Required access depends on the selected signal. Status: proposed uncertainty-dependent application. |
| **A13 Agent Action Ribbon** | Align tool calls, returns, retries, and execution status with the relevant answer segment. AgentLens and AGDebugger provide agent-behavior and debugging precedents [37], [15]. An action log establishes recorded execution, not the correctness of a resulting conclusion. Required access is timestamped tool events. Status: existing research interface precedents. |
| **A14 Provenance Descent** | Navigate from an answer to source-linked reasoning, external evidence, and any available internal observations. Graphologue and details-on-demand provide linked-inspection precedents [26], [48]. At each step, the interface should identify the source and its evidential status. Required access is provenance plus independently available lower-layer channels. Status: partial precedents, proposed traversal across evidence types. |
| **A15 Second-Opinion Overlay** | Place another model's assessment beside the selected claim and expose agreement, disagreement, and supporting evidence. Multiagent debate and LLM Comparator provide methodological and interface precedents [14], [27]. Additional calls incur latency; agreement among related models is not independent confirmation. Required access is separately recorded evaluations. Status: partial precedents. |

## Appendix B Application Stack and Reasoning Access

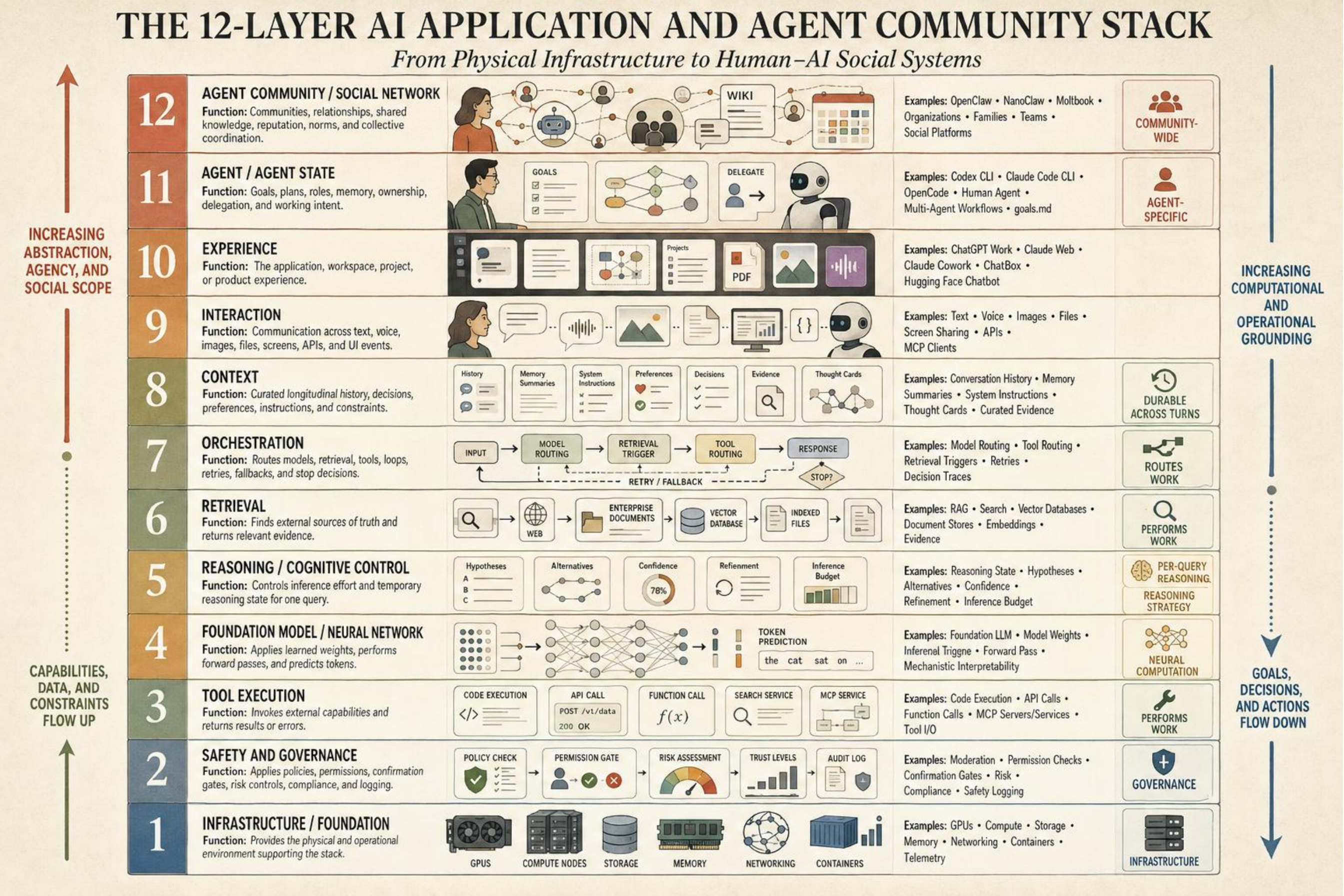

**Figure 4. The current application stack is information rich and well suited to support visualization and sensemaking across all levels. The Reasoning Workspace can leverage observations from application context, orchestration, model execution, and infrastructure. Knowledge is synthesized across the stack, not solely from the neural network.**

Figure 4 outlines the workspace capability within the application stack. The stack is a conceptual organization of responsibilities. Context, orchestration, retrieval, tool activity, and model diagnostics introduce different observations. Governance and infrastructure constrain which measurements can be captured and retained.

The measurement record tracks the target checkpoint, run, input prefix, capture location, token positions, and analysis procedure. An activation decoder leverages its own checkpoint and question. An intervention needs a baseline, intervention target, and outcome metric. A summary needs a link to the text it summarizes and the method that produced it. Without these distinctions, an interface cannot reliably align observations or reproduce an explanation.

### B1 Evidence channels and their limits

Generated reasoning gives access to an emitted sequence [57], [53], [31]. Sparse features and attribution graphs expose a selected representation or approximation [1], [11]. Intervention results test effects within a specified experimental contrast [9]. Activation-conditioned natural-language reports describe information accessible to a decoder [24], [43], [28], [47], [22]. Introspective self-report requires separate validation; controlled concept-injection experiments offer evidence under particular conditions rather than general access to an internal narrative [34]. The Jacobian lens bridges vocabulary-oriented readout and intervention analysis [25].

Residual-state storage across L layers, T positions, and width d costs $L \times T \times d \times b$ bytes at b bytes per element, excluding weights and other runtime state. Selective capture reduces cost but changes the available evidence; the API must identify capture scope, truncation, and unsupported measurements.

## B2 Interoperable interface requirements

The proposed requirements extend OpenAI-compatible interfaces while recognizing that the native OpenAI Responses API has its own representation [42], [38], [12], [59]. We do not prescribe a new field schema. A capability description should tell the client which reasoning, tool, statistical, and internal channels exist; whether they are streamable; whether they can be replayed; and whether an intervention requires a new run. Availability should be specific to the model, endpoint, and request mode.

The client should preserve returned evidence and construct a separate context representation for each human-approved branch. Opaque continuation data cannot be edited as readable reasoning. If an exclusion conflicts with an opaque state or a provider's preservation requirements, the branch must restart from supported context rather than claim that the unwanted material was removed from hidden state. Access control, retention, and redaction requirements apply to both readable and encrypted records [44].

NNsight/NDIF demonstrates remote observation and intervention on supported open-weight models [21]. SGLang exposes optional hidden-state and routed-expert outputs in supported configurations [46]. Neuronpedia's interpretability infrastructure provides feature inspection and related tools [39]. These systems provide starting points for interoperability experiments. Whether their evidence channels help ordinary users remains to be tested.

## Acknowledgements

This research is executed by the Virginia Tech Lab for Information Visualization (http://infovis.cs.vt.edu/) and sponsored by Peraton Labs. Special thanks go to Mazal Bethany and Greg Caufman for their support! Additional thanks go to Rob Capra, Wendy Schafer-Colvin and Dave Korb for their support!